\documentclass[12pt]{iopart}   

\usepackage{iopams}
\usepackage{graphicx}   
\usepackage[basic]{circ}

\begin{document}

\title{Matter and Interactions: a particle physics perspective}
\author{Giovanni Organtini}
 \address{"Sapienza", Universit\`a di Roma, P.le A. Moro 2, ROMA I--00185 Italy}
 \ead{giovanni.organtini@uniroma1.it}   
\date{\today}

\begin{abstract}
In classical mechanics matter and fields are completely separated. Matter interacts with fields. For particle physicists this is not the case. Both matter and fields are represented by particles. Fundamental interactions are mediated by particles exchanged between matter particles. In this paper we explain why particle physicists believe in such a picture, introducing the technique of Feynman diagrams starting from very basic and popular analogies with classical mechanics, making the physics of elementary particles comprehensible even to high school students, the only prerequisite being the knowledge of the conservation of mechanical energy.
\end{abstract}

\maketitle

\section{Introduction}
In this lecture I try to make popular particle physics, including those mathematical details usually completely skipped. The lecture is aimed to degree students of disciplines other than physics, such as engineering or informatics, but it was proven to be effective for medicine students, too. With some effort, it can even be understood by high school students. The prerequisites, in fact, are just the knowledge of the concept of energy, the principle of its conservation and some basics of mathematics. In the lecture, paragraphs written with smaller characters can be skipped by those with limited knowledge of mathematics.

\section{Particles and Fields}
Particle Physics is a very ambitious field. It aims to describe of which the Universe is made as well as how its most fundamental constituents interact between them. In other words, it is the science of elementary particles and fundamental forces. Because of its name, particle physics is often thought to be related to matter only. The fact that matter is made of particles, in fact, is widely accepted. However, forces are the subject of particle physics, too, because of quantum mechanics. Quantum mechanics predicts that matter behaves like waves or fields under certain circumstances, as well as fields and waves behave as particles. Matter and fields, then, are tightly bound each other in quantum mechanics. 

Which forces are investigated in particle physics? Certainly not elastic forces, friction, chemical bounds, etc. All these forces are {\em non fundamental}. On the contrary, gravity, electromagnetism, weak and strong forces are thought to be {\em fundamental}. What makes a force fundamental is the fact that one can describe the interaction in terms of fields. What is a field?

It is well known that Sun attracts Earth by a gravitational force $F_{se}=m_e {\cal G}$ proportional to the mass $m_e$ of the Earth. The Sun attracts even Mars with a gravitational force $F_{sm} = m_m {\cal G}$ proportional to the mass $m_m$ of Mars. All the bodies in the solar system are attracted by the Sun (and by all other bodies) with a force proportional to their mass. The {\em constant} ${\cal G}$ depends only on the Sun properties and on the distance with respect to it. We believe that the existence of the field is due just to the fact that Sun exists; ${\cal G}$ would exist even if there is no Earth or Mars to show us that a force develops between those two bodies. We call ${\cal G}$ a {\em field}. A force develops when particles fall within a field. One can imagine a field as a modification of the properties of the space around a given body, called the {\em source} of such a field, whose effect is to produce forces proportional to its strength when other sources are found in it.

We say that mass is the source of the gravitational field. Electric charge is the source of the electric field. Currents are the sources of the magnetic field. Each of these sources provides a field, i.e. modify the space around them in such a way that, putting another source of the field in that space produces a force between them. For example, electric charges generate electric fields. Putting another electric charge in such a field produces a force. No force is expected putting in the electrostatic field a neutral body, since such a body is not in turn a source of the electric field. A magnetic field develops around an electric current flowing in a conductor. Currents are moving electric charges. If we put any body within this field, it does not show any force unless it has an electric charge and has non null speed. 

Fields need a source. While for gravity the source is mass, there is no source for friction, nor for elastic forces. Today we know that they are manifestations of fundamental forces in complex systems. Elastic forces, for example, are nothing but electromagnetic forces. The atoms of a spring are bounded together because of Coulomb--like forces. There is an equilibrium between repulsive forces and attracting ones. If you try to extend the spring, the atoms become farer each other and at some time attractive forces dominate, causing the spring to develop a force who is, in fact, the sum of elementary, individual electrostatic forces.

While gravity and electromagnetism are widely known to general public, this is not true for strong and weak forces. The strong force is the one who binds together protons in the nucleus of atoms, for example. Electrostatic forces tend to disrupt a nucleus composed of more than one proton. If two or more protons are bound together, there must be another, attractive force. Such a force cannot be gravity. It is too weak to keep the protons in the nucleus. 

\begin{center}
\begin{minipage}{0.9\textwidth}
\footnotesize
To realize how weak is the gravitational force, let's compute its intensity and let's compare it to the electrostatic repulsion between two protons at a distance of the order of the diameter of a nucleus. The size of atomic nuclei is of the order of $1-2$~fm, i.e. $r \simeq 10^{-15}$~m. The gravitational force between two protons ($m_p \simeq 1.7\times10^{-27}$~kg) at such a distance is, then

\begin{equation*}
F_g = G\frac{m_p^2}{r^2}\simeq 6.67\times10^{-11}\frac{2.9\times10^{-54}}{10^{-30}}\,\mathrm{N}\simeq 19.3\times 10^{-35}\,\mathrm{N}
\end{equation*}
The electrostatic force, being $e=1.6\times10^{-19}$~C the electric charge of the proton, is

\begin{equation*}
F_e = k\frac{e^2}{r^2}\simeq 9\times10^9\frac{2.6\times 10^{-38}}{10^{-30}}\,\mathrm{N}\simeq 23.4\,\mathrm{N}
\end{equation*}
The electrostatic repulsion is about 35 orders of magnitude stronger than gravitation. Gravity, then, cannot be responsible for binding protons together in the nucleus, even if they are close to each other.
\end{minipage}
\end{center}

Weak force is unknown to the majority of people. It is usually said that it is responsible for nuclear decays. A nuclear decay consists in the transmutation of an atom in another, lighter atom, accompanied by the emission of some radiation like photons (also called $\gamma$ radiation), electrons ($\beta$ radiation) or Helium nuclei ($\alpha$ radiation). Forces are needed to modify the status of a particle. In classical mechanics, the status of a particle is given by its position and its velocity. Changing the velocity of a particle implies acceleration, i.e. a force. In quantum mechanics, the status of a particle is given by its mass, energy and angular momentum. For a particle to change its status a force is required. In a nuclear decay, the mass and/or the energy of a particle changes, then a force is required. That is why we need a weak force, responsible for neutrino interactions, too, to explain $\beta$ radiative decays.

\section{Dynamics}
The equation of motion of particles in a field, whatever their nature, can be written as simple as 

\begin{equation*}
\Delta U = 0\,.
\end{equation*}
That is quite trivial, in fact. You can describe the whole Universe by such an equation. Any physics law can be written as $f(x)=0$. For example, you can write $F=ma$ as $f_1=F-ma=0$, or $V=RI$ as $f_2=V-RI=0$. Then, summing up all the known physics laws you can define $\Delta U=\sum{f_i}=0$. That is not very useful. On the contrary, if you are able to give a simple enough definition for $U$, you may find that all the dynamics of particles can be described by what is called a {\em variational principle}. The equation above states, in fact, that $U$ is a constant. A variation in the value of $U$ in an isolated system (i.e. a system that cannot exchange matter nor energy with other systems) must be null. From this principle comes the laws of physics.

You should already know at least an example of this principle. Consider a mechanical system composed of a particle of mass $m$ in a uniform gravitational field with acceleration $g$. The particle energy $U$ is the sum of its potential energy $V=mgh$ and its kinetic energy $K=\frac{1}{2}mv^2$. The height $h$ of the particle is measured with respect to a conventional level, for which we define $U=0$. The energy of the particle is conserved, i.e. $U=const$ at any time. That, of course, does not mean that nothing changes. In fact, if you leave the particle falling down, its speed increases, while its position varies with time. The variational principle states that a variation of $U$ must be null. Varying $U$ corresponds to taking its derivative and put it equal to zero.

\begin{center}
\begin{minipage}{0.9\textwidth}
\footnotesize
Let's take this derivative and impose that it must vanish. Since $U$ is a function of both $h$ and $v$, the variation $dU$ of $U$ is

\begin{equation*}
dU = \frac{\partial U}{\partial h}dh + \frac{\partial U}{\partial v}dv=mg\,dh + mv\, dv\,.
\end{equation*}
If $dU=0$, then $dU/dt=0$, and

\begin{equation*}
\frac{dU}{dt}=0=mg\frac{dh}{dt}+mv\frac{dv}{dt} = mgv+mva\,.
\end{equation*}
where $a=dv/dt$ is the acceleration of the particle.
\end{minipage}
\end{center}
For those who do not know the concept of derivative, we can show that in fact the result is the one given above as follows.

If $\Delta U = 0$, then $\frac{\Delta U}{\Delta t}=\frac{\Delta V}{\Delta t}+\frac{\Delta K}{\Delta t}=0$ where $\Delta t$ is the time needed to make $U$ vary by $\Delta U=\Delta V + \Delta K$. Let's compute the variation $\Delta U$ in a given time, considering the two terms separately. The first term is $V=mgh$. Since $m$ and $g$ are constants, only $h$ can change and $\Delta V = mg \Delta h$. As a result 

\begin{equation*}
\frac{\Delta V}{\Delta t} = mg\frac{\Delta h}{\Delta t}=mgv
\end{equation*}
For the second term we have $\Delta K = \frac{1}{2}m\left(v(t+\Delta t)^2-v(t)^2\right)$, where $v(t)$ is the speed of the particle at time $t$. Of course $v(t+\Delta t) = v(t) + \Delta v$, i.e. the speed at time $t+\Delta t$ is the speed at time $t$ plus some speed $\Delta v$. We then have
\begin{equation*}
v(t+\Delta t)^2 = v(t)^2 + \left(\Delta v\right)^2 + 2v(t)\Delta v
\end{equation*}
and, substituting in the expression for $\Delta K$ we get

\begin{equation*}
\frac{\Delta K}{\Delta t} = \frac{1}{2}m\left(\frac{v(t)^2}{\Delta t} + \frac{\left(\Delta v\right)^2}{\Delta t} +
2v(t)\frac{\Delta v}{\Delta t} - \frac{v(t)^2}{\Delta t}\right)
\end{equation*}
The first and last terms in the parenthesis cancel each other, while, choosing $\Delta t$ small enough, $(\Delta v)^2$ can be made much smaller than $\Delta v$ and can be neglected, so that

\begin{equation*}
\frac{\Delta K}{\Delta t} = \frac{1}{2}m\left(2v(t)\frac{\Delta v}{\Delta t} \right)=mv\frac{\Delta v}{\Delta t}=mva
\end{equation*}

In the end, we have $mgv+mva=0$, i.e. $-mg=F=ma$, the Newton equation. The dynamics of particles in uniform gravitational fields, then, comes from a variational principle. The fundamental law is the one expressing energy conservation. Newton's law is a consequence of energy conservation.

The same is true in non--uniform gravitational fields. In fact, the definition $V=mgh$ is just an approximation. We all know that gravity is not uniform, but scales as $1/h^2$. However, if $h \ll r_e$, where $r_e$ is the earth radius, it can be considered as uniform. This is just a result of a general technique known as the Taylor expansion. According to this technique, any function $f(x)$ (with some {\em good} properties, like continuity, derivability, etc.) can be written as a series, involving the derivatives of the function being expanded.

Again, we can give an intuitive picture of what Taylor expansion is to those who don't know about derivatives. Consider, for example, the temperature at a given distance $x$ from a heater. The temperature $T$ depends on the distance. The higher the distance, the lower the temperature. We say that $T$ is a function of $x$ and write $T$ as $T(x)$.

We can assume that the temperature at the surface of the heater (i.e. at distance $x=0$ from it) is constant $T(0)=T_0$. As far as the distance is short, the temperature $T$ can be considered constant and equal to $T_0$ at any distance $x>0$ small enough, so that $T(x)\simeq T_0$.

If $x$ becomes larger, however, this is no more true. We know that the temperature decreases. As a first approximation we can write $T(x)$ as $T(x)=T_0 - \alpha x$, so that the value of $T$ decreases with $x$. However this behavior can be considered close to the real behavior only for small $x$. As $x$ increases further, we may find that the temperature decreases too fast. In fact there will be a point for which $T(x)=0$, when $x=T_0/\alpha$, that is unreasonable. A possible way to reduce the {\em speed} at which $T$ decreases is to add an extra term that, however, must not introduce a dramatic change when $x$ is small. If the extra term is proportional to $x^2$, when $x$ is small, $x^2$ is even smaller and does not contribute sensibly to the value of $T(x)$. We can write then

\begin{equation*}
T(x) \simeq T_0 -\alpha x + \beta x^2
\end{equation*}
The extra term protects $T(x)$ from becoming too small, but only to a given extent. There is still some value of $x$ for which $T(x)=0$. The reason being that $T(x)$ cannot probably be written as a polynomial. However, polynomials are good enough representation of the behavior of the temperature as far as the distance between the point at which we know its value with infinite precision is small.

That is Taylor expansion: a way to express an unknown function in terms of a sum of powers of the variable on which the function depends upon. 

\begin{center}
\begin{minipage}{0.9\textwidth}
\footnotesize
For those who knows about derivatives, we can write the Taylor expansion more formally. It can be shown that any {\em regular} function can be written, as

\begin{equation*}
f(x) \simeq f(x_0)-\left.\frac{df(x)}{dx}\right|_{x=x_0}(x-x_0)+\left.\frac{1}{2}\frac{d^2f(x)}{dt^2}\right|_{x=x_0}(x-x_0)^2+\cdots
\end{equation*}
for $x$ close enough to $x_0$. 
\end{minipage}
\end{center}

The exact form of the gravitational potential $V$ is

\begin{equation*}
V=-G\frac{M}{r}\,,
\end{equation*}
and it can be shown that its Taylor expansion (as you may easily compute, knowing the concept of derivative) is

\begin{equation*}
V \simeq -G\frac{M}{r_e}-G\frac{M}{r_e}\left(\frac{r-r_e}{r_e}\right)-G\frac{M}{r_e}\left(\frac{r-r_e}{r_e}\right)^2+\cdots
\end{equation*}
The first term corresponds to the constant term that we usually put equal to zero, since adding or subtracting constants to potential energy can be done without affecting the results (remember that the derivative, i.e. the variation of a constant is zero). The second term of the expansion is just the term $mgh$. In fact $g=GM/r_e^2$, while $h=r-r_e$. To write the potential energy we just multiply this term by the constant $m$.

\section{The equation of the Universe}
Given the ingredients outlined above, we can now proceed in writing the fundamental equation for the particles and fields of which the Universe is made. Remember that, in quantum mechanics, particles and fields are almost the same object. From the mathematical point of view they are both {\em fields}, where this term denotes a complex object who obeys a non conventional algebra, but, nevertheless, can be thought mostly as a number in our very simple treatment.

Our {\em equation of the Universe} is always of the form $\Delta U = 0$, where $U=K+V$ is now a function of the fields. The simplest expression for $U$ is just a product of all the independent variables determining the energy of the Universe, such that the physical dimensions of the product are those of the energy. There must be at least two particles to give rise to some interaction. Let's call $\bar{\psi}$ and $\psi$ the fields of those two particles. They are some function of the state of each particle. They depend, then, on mass, position, momentum, angular momentum, and any other number characterizing the particles.

If we call $A$ the field representing the interaction, a possible expression for the interaction potential $V$ is $V=\alpha \bar{\psi}A\psi$, where $\alpha$ is some coupling constant, defining the {\em strength} of the interaction. The higher $\alpha$, the stronger the interaction.


The term $\alpha \bar{\psi}A\psi$ must be analogous to the first approximation of the potential energy in the case of gravitational field, $mgh$. Then it can be conceived as a term of the Taylor expansion of the function $V$ representing the energy of the Universe:

\begin{equation*}
V \simeq V_0\left( 1 - \alpha\bar{\psi}A\psi + \frac{1}{2}\alpha^2\left(\bar{\psi} A \psi\right)^2+\cdots\right)\,.
\end{equation*}
The constant term is irrelevant. What is important for dynamics is in fact the variation of $V$, so that the constant will vanish, once the derivative of $V$ is taken. Despite the fact that the product $\bar{\psi}A\psi$ is not an ordinary product between numbers, a theorem known as Wick's theorem~\cite{wick}, provides simple rules about how to compute the various terms of $V$. The theorem states that the product must be taken in such a way that terms of order $\alpha^n$ must  be obtained evaluating the fields in $n$ different points. It is due to Richard Feynman~\cite{feynman} the observation that each term in the Wick's theorem can be graphically represented in such a way that identifying the relevant terms for a given process is as simple as writing lines on a piece of paper forming connected graphs. For this reason, these graphs are called {\em Feynman diagrams}.

The rules to write a Feynman diagram are as follows: when you read a $\bar{\psi}$ write an arrow, entering in one point of the space--time; an arrow must be written also for $\psi$, but it has the opposite direction, i.e. it comes out of that point. $A$ translates into a wavy  line. The points in which all the lines comes together are those foreseen by the Wick's theorem and are called {\em vertices}. Of course, real Feynman rules are much more complex than the ones we are illustrating here. They were oversimplified to make them comprehensible to most of you. What is important, here, is the transmission of the concept behind them. 

\begin{figure}
\begin{center}
\includegraphics[height=0.2\textheight]{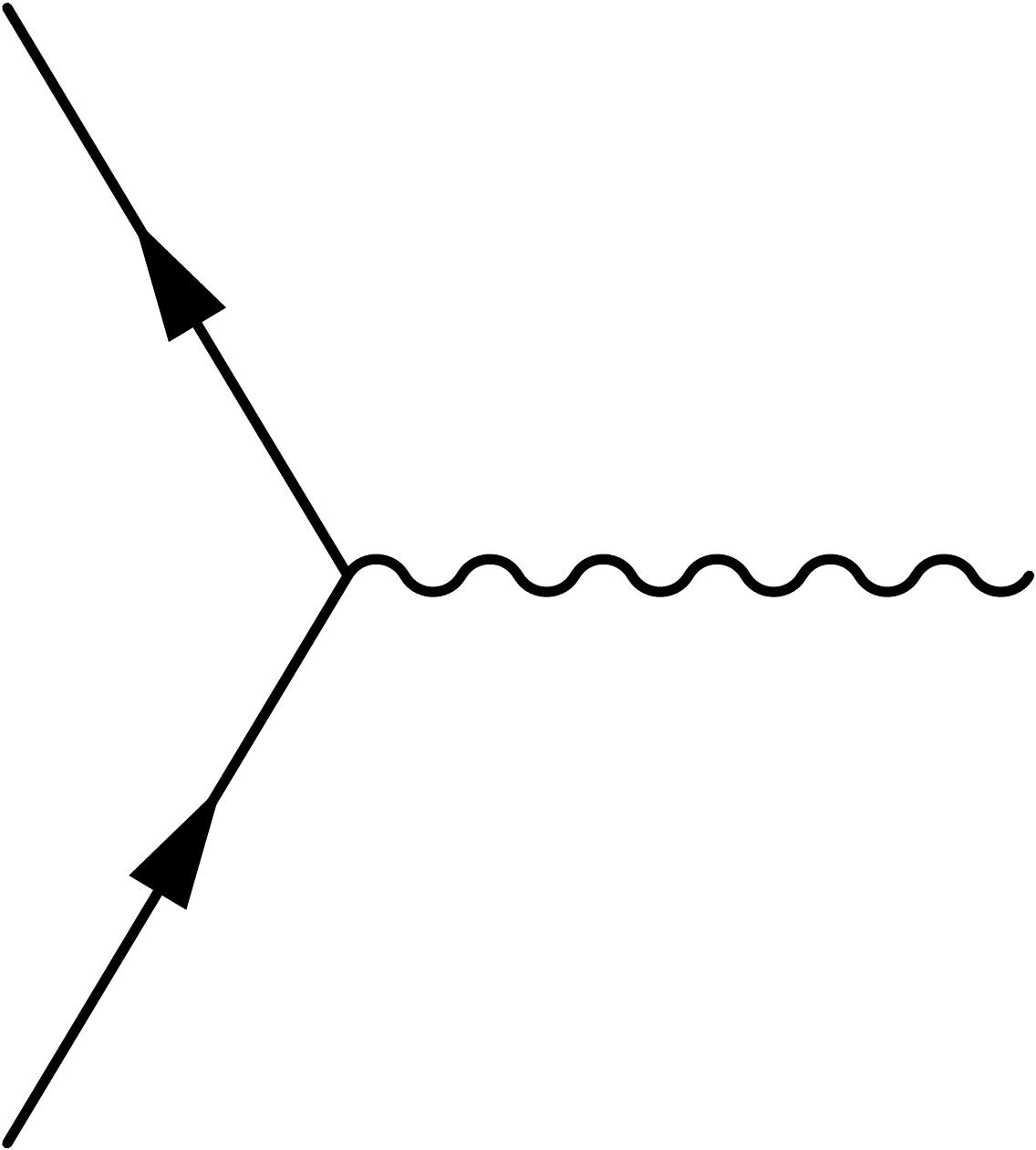}
\caption{\label{fig:treelevel}A very simple Feynman diagram showing the interaction between two particles and a field.}
\end{center}
\end{figure}

The first term of the expansion, then, translates in a diagram like the one in Figure~\ref{fig:treelevel}.

If $\psi$ represents, for example, an electron, and $A$ another particle, $\bar{\psi}$ is an electron, too, moving in the opposite direction with respect to the vertex (the point in which the three lines merge together). There is just one vertex for this term, since it is of order $\alpha$ ($n=1$). The Feynman diagram in Figure~\ref{fig:treelevel} can be thought as the visual representation of a process in which an electron coming from bottom left, emits a particle represented by $A$ and, consequentially, modify its direction toward top left, to conserve momentum. It is important to understand that this is not necessarily what happens at microscopic level. The {\em reality} is described by the equation of motion, not by the Feynman diagram. The latter is just another, funny way to write $\alpha\bar{\psi}A\psi$. What is important for physics are the predictions made by equations. As far as these predictions conform to experimental results, we are then free to interpret the diagram as exactly what happens at microscopic level.

Quantum mechanics allows the computation of the probability of a given process. Computing the first term of the $\Delta V$ expansion (the one given by $\alpha\bar{\psi}A\psi$) gives zero, i.e. such a process is forbidden. In fact, there is no way for an electron to undergo such a process, since it does not conserve energy and momentum. That is encouraging, in fact.

Since the first term of the expansion gives null results, the first non--trivial term must be the second one, i.e. the one proportional to $\alpha^2$.

\begin{figure}
\begin{center}
\includegraphics[height=0.2\textheight]{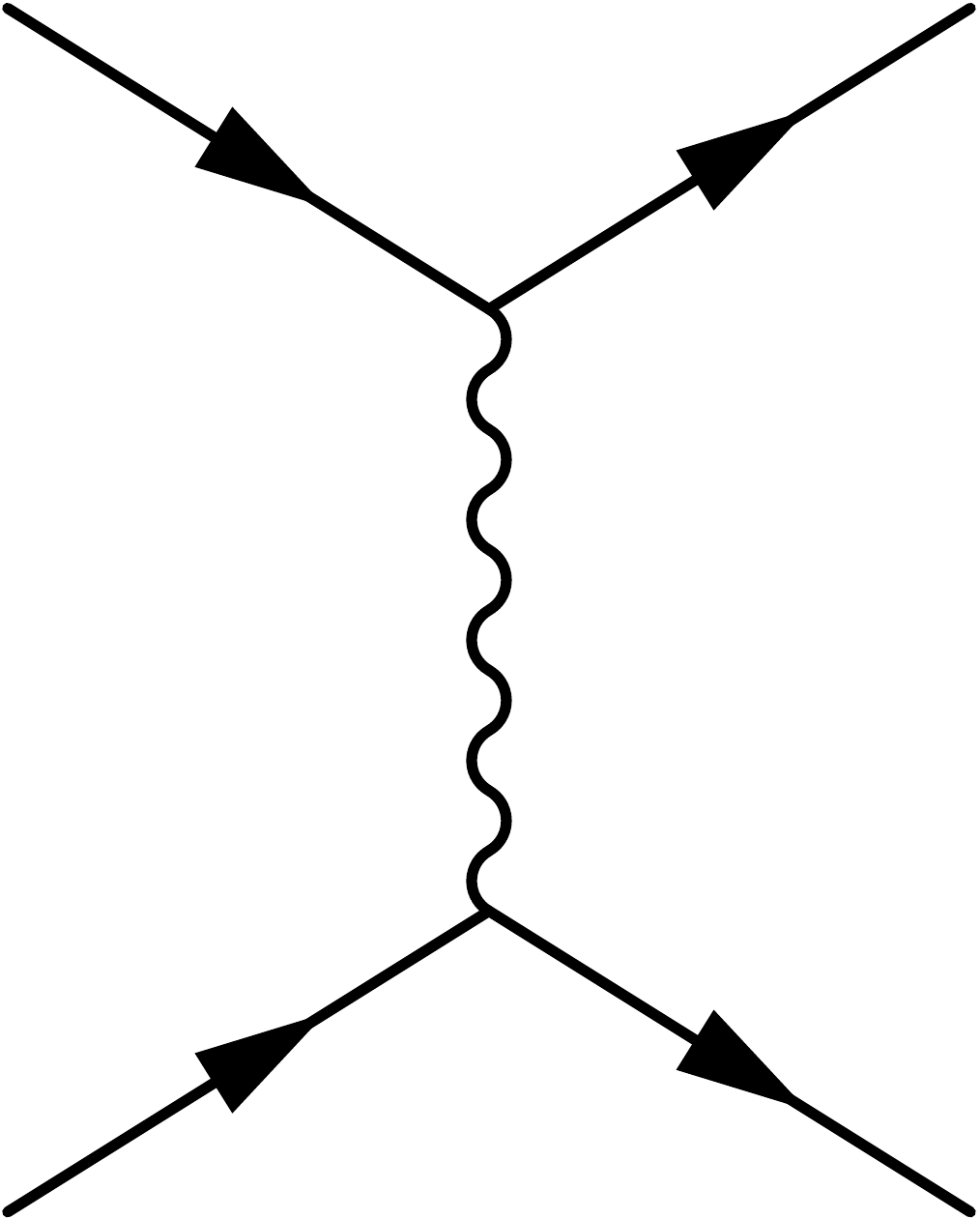}
\caption{\label{fig:eeee}A Feynman diagram generated by taking the second term of the Taylor expansion of the variation of the energy.}
\end{center}
\end{figure}

To build the corresponding Feynman diagram, one can take two Feynman diagrams like those in Figure~\ref{fig:treelevel} and join them together. In this way we have two vertices, as prescribed by the Wick's theorem. A possible way is joining the two curly lines, by which one obtains the diagram in Figure~\ref{fig:eeee}.

The associated term in the expansion corresponds to two electrons in the initial state, interacting via the electromagnetic field represented by $A$. The final state is composed of two electrons, too. That's why there are four electron fields (two for the initial state and two for the final state).  This term must give the probability of interaction of two electrons, then.
Computing this term we see that it is non zero and it is incredibly similar to the expectations from experiments. 

We can interpret this diagram as follows: two electrons (one at the bottom and one at the top of the figure) come close to each other. One emits a particle that we call a {\em photon}, who is absorbed by the other. Both, then, change their momentum and proceed in their motion. The net result is the scattering of two electrons, initially coming close to each other, then getting apart. In other words, we are describing the electrostatic repulsion between the two electrons in terms of the exchange of a photon. With this respect, the photon can be seen as the {\em mediator} of the electromagnetic field. The electrostatic force is nothing but the result of the exchange of a photon, that is not visible in the final state, but {\em carries} the electric field.

That's why particle physics include particles and fields as the same object. Both are particles: some of them are matter particles, some other are force mediator particles. We do not use Feynman diagram because we believe {\em a priori} that particles interact between them exchanging particles, so we can write the interaction in terms of products of the corresponding fields. It is just the opposite: since the terms who enters in the equation of motion can be graphically represented by a Feynman diagram and results are compatible with experiments, we are free to interpret Feynman diagrams as what happens at microscopic level and we are lead to the conclusion that interactions can be represented by the exchange of mediators.

\begin{figure}
\begin{center}
\includegraphics[height=0.15\textheight]{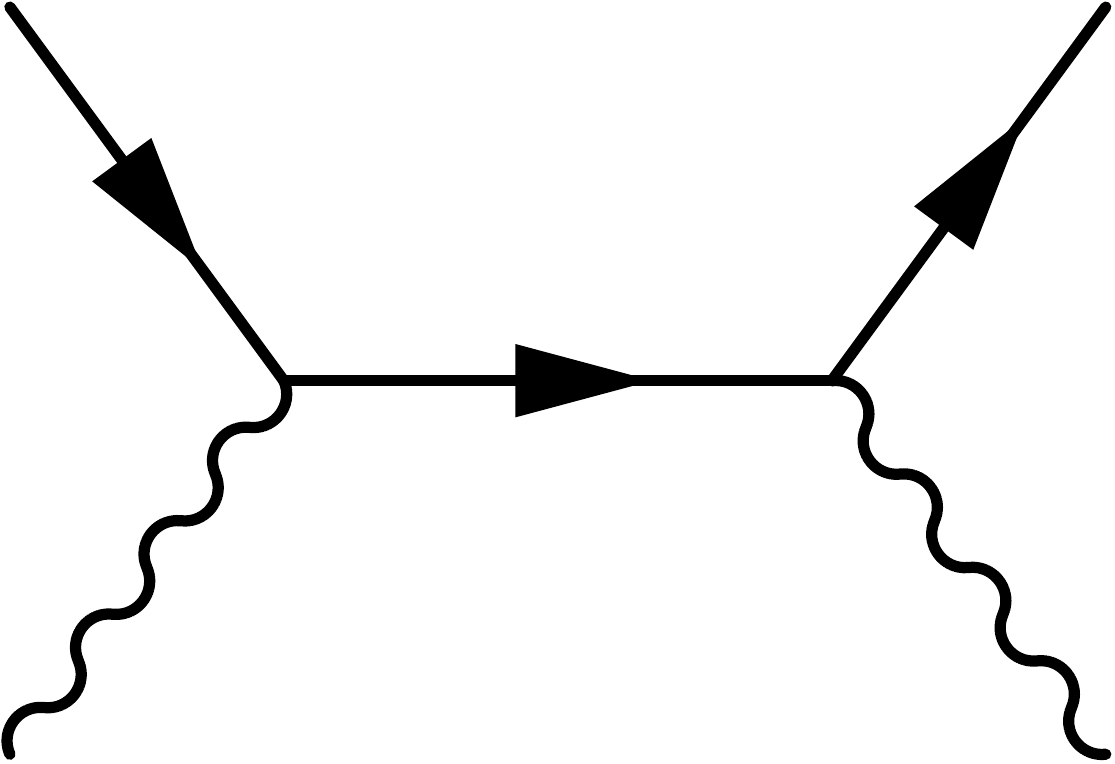}
\caption{\label{fig:einteraction}Another Feynman diagram generated by taking the second term of the Taylor expansion of the energy variation.}
\end{center}
\end{figure}

Another possibility is to write a Feynman diagram composed of the two terms of order $\alpha$ joined by the particle leg, instead of the photon, like in Figure~\ref{fig:einteraction}.

This diagram is a valid diagram and its computation gives rise to a non--vanishing probability. In fact such a diagram can be interpreted as the interaction of an electron in an external electromagnetic field. The field is modeled by the photons. In both the initial and final state there are one electron and one photon. This same diagram represents the Compton scattering, where a photon interact with a free electron and emerges with a different wavelength, while the electron acquire some energy.

In summary, thanks to Feynman diagrams, we can represent visually an equation and, in turn, we can interpret these diagrams as what really happens microscopically. In the end, we can interpret the fields as particles acting as force mediators. Matter particles exchange mediators between them to interact. The interaction between two particles, in fact, is described in terms of a mediator sent from one particle to another and vice versa. Just as two tennis players remain within the field while they are playing, exchanging the ball between them, attractive forces are modeled in such a way that two particles need to be close to each other to continue {\em playing} tennis using a ball made by a force mediator particle. As far as players play, they are bound to each other. Once the match is over, no particle is exchanged between the players and they are free to move away. Repulsive forces can be thought as a shooter who shoot a disk with a gun. The shooter receives a hit from the gun, who expels the projectile (our force mediator) that, in turn, transfers momentum to the disk. The disk and the shooter, then, move away from each other.

\section{Higher order corrections}
As said in the previous section, the computation of the probability for a process to happen, according to the theory outlined above (called {\em quantum electrodynamics} or {\em QED}), is very close to that experimentally measured.

The predictions from QED differs from experiments only few percent. However, this is not the end of the story. In fact, in the computation we just took the first non vanishing term of the expansion. We can write other diagrams, with a higher number of vertices, corresponding to higher terms in the expansion.

It is well known, even from classical physics, that when electrons accelerate, produce an electromagnetic field. In terms of QED the field is modeled by photons. In electron scattering, represented by the diagram in Fig.~\ref{fig:eeee}, at least one electron may be accelerated enough to produce an electromagnetic field, i.e. a photon.

\begin{figure}
\begin{center}
\includegraphics[height=0.15\textheight]{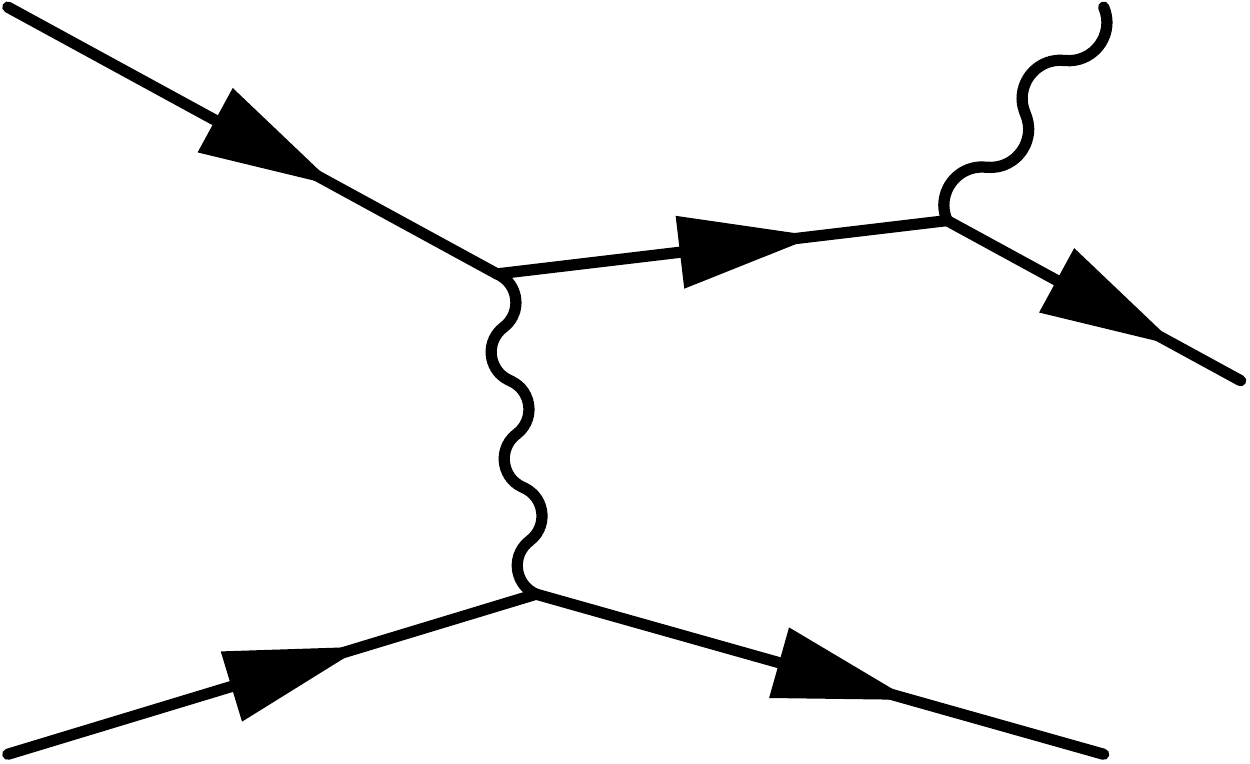}
\caption{\label{fig:emission}The Feynman diagram representing the third term in the expansion of the e.m. interaction potential.}
\end{center}
\end{figure}

The next term in the expansion of $V$ reproduces exactly this situation, as can be seen from Figure~\ref{fig:emission}. This diagram, as expected, has three vertices, and can be built joining three basic diagrams, like those in Fig.~\ref{fig:treelevel}. It represents a process where two electrons come close to each other, interact and are scattered away, with the emission of some electromagnetic field.

\begin{figure}
\begin{center}
\includegraphics[height=0.15\textheight]{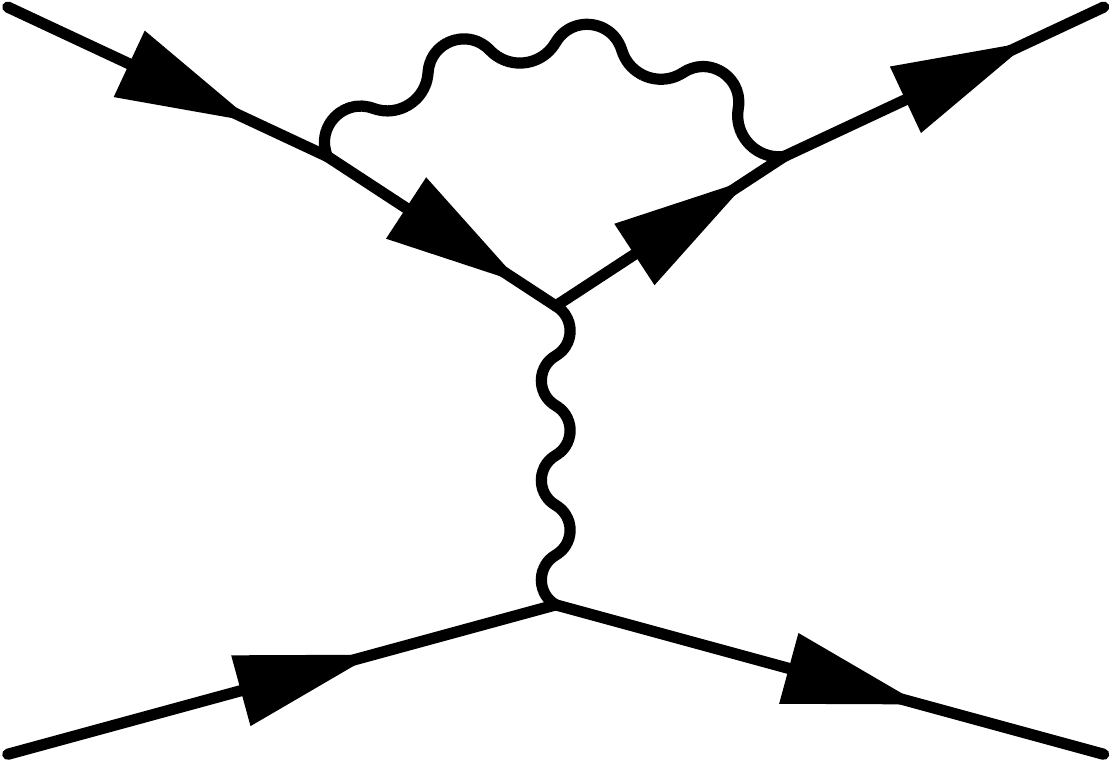}
\caption{\label{fig:loop}A higher order term in the expansion for the process of electron scattering.}
\end{center}
\end{figure}

Moreover, there are processes modeled by higher order terms, for which the final state is still composed of two electrons. In evaluating the probability for electron scattering, we must take into account these terms, since they are part of the expansion. One of these terms is represented by the Feynman diagram shown in Fig.~\ref{fig:loop}, which has four vertices, representing a term of order $\alpha^4$, and is obtained joining four basic diagrams.

The associated probability is small (smaller than the one associated to the main diagram). However, adding the corresponding probability to the one obtained computing the diagram in Fig.~\ref{fig:eeee} gives a result much closer to the experimental value. If we add more terms we obtain results closer and closer to those experimentally measured. We are able to compute these kind of processes with a precision up to few permille. No other process in physics is known to this level of accuracy.

Another, very nice and accurate, yet popular, description of the Feynman diagrams technique is given in~\cite{daniel}.

\section{Antiparticles}
At the beginning of XX century, P.A.M.~Dirac~\cite{dirac} noticed that the equation of motion of free electrons moving back in time, can be regarded as the equation of motion of a free positron, i.e. a free electron with positive electric charge. Positrons are called the antiparticles of the electrons. Generally speaking antiparticles are just ordinary particles with opposite charges. Then, antiprotons are negatively charged protons.

Positrons where in fact discovered in cosmic rays in 1932 by C.~Anderson~\cite{anderson}. However, positrons are not usually found in ordinary matter. So, how is it possible to observe them?

\begin{figure}
\begin{center}
\includegraphics[height=0.15\textheight]{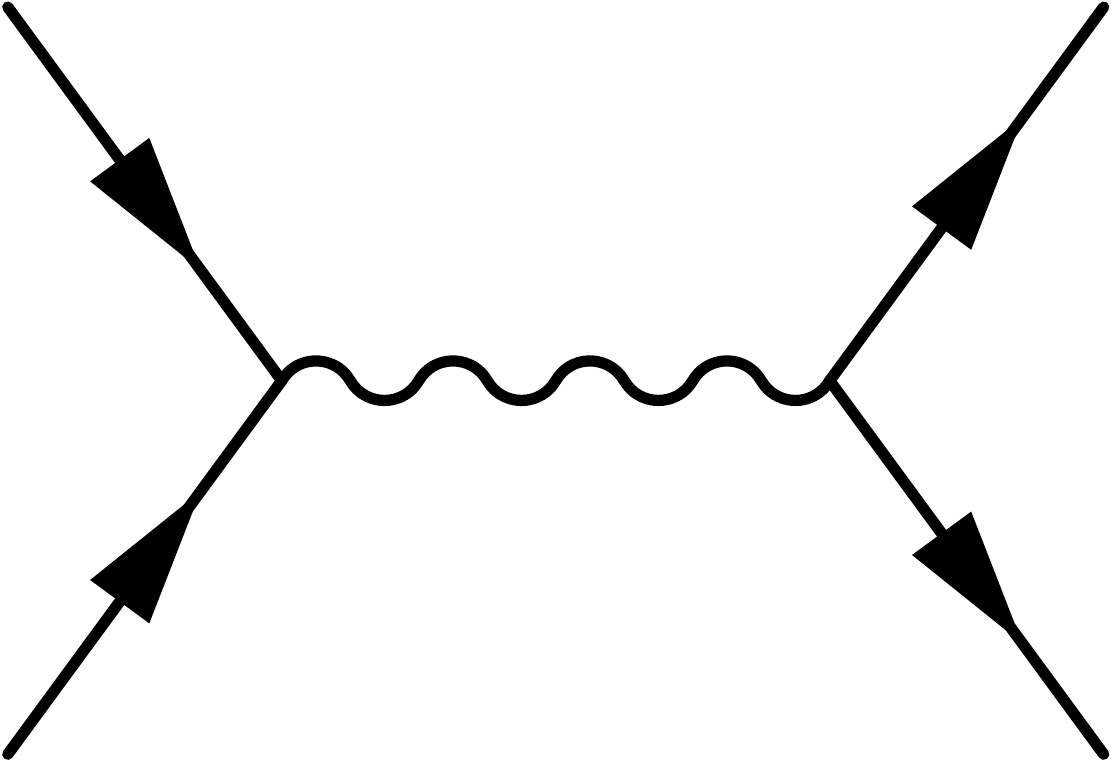}
\caption{\label{fig:annihilation}A Feynman diagram for annihilation of particles and antiparticles.}
\end{center}
\end{figure}

Feynman diagrams, together with the observation from Dirac, give us the answer. If we take the diagram in Figure~\ref{fig:eeee} and rotate it clockwise by 90~degrees, assuming time flowing from left to right, we notice that there are two particles moving back in time: one in the initial state and one in the final state. They can be interpreted, according to Dirac, as antiparticles moving forward, as in the diagram shown in Figure~\ref{fig:annihilation}.

In this diagram the two legs moving from left toward the vertex are a particle and its antiparticle. When they comes together they are said to {\em annihilate} into a photon. The photon, then, is said to {\em materialize} in a particle--antiparticle pair. In fact, photons with enough energy, whatever they are produced, can materialize in particle--antiparticle pairs. This explains how it is possible to observe antiparticles: they are the products of materialization, described by our theory that, in turn, can be outlined in terms of Feynman diagrams.

Needless to say that the probability for this process is predicted by the theory and is found to be exactly the size of the experimentally observed probability.

By the way, this is also the process used in colliders to produce new particles. There are currently two large colliders running in the world. One is the Fermilab's Tevatron, close to Chicago, USA; the other is CERN's LHC, at the border between France and Switzerland. The first makes protons collide with antiprotons; the latter makes protons collide with other protons. We know that protons are not elementary particles. They are composed by quarks and antiquarks. In both accelerators, quarks annihilate with antiquarks and produce other massive particles. The energies of the colliders are large enough to give the photons enough energy to materialize in a couple of massive particles. So massive, in fact, that they possibly cannot be produced in ordinary conditions. Such an energy must be of the order of $E\simeq m c^2$, as predicted by Einstein.

\section{Other interactions}
Using Feynman diagrams we explained the phenomenology of electromagnetic interactions, identifying the photon as the electromagnetic force mediator. It turns out that the same theory can be made for other interactions. For example, if $A$ in the expansion terms of the energy variation is taken to be the field of strong or weak fields, and $\alpha$ is replaced by the corresponding coupling constants, the same technique can be applied and we can write down Feynman diagrams for which the curly lines represent the mediators of those forces rather than a photon. The diagrams appear exactly the same, but the {\em values} of the fields are different and the computation of the probability differs in size, but not in principle. 

Other mediators, then, should exist for weak and strong forces, as well we believe the same applies for gravitation. In fact, using colliders, we can produce such mediators and observe their decay products, i.e. the particle--antiparticle pairs in which they materialize. 

In 1983 the CERN's experiment~\cite{SpS}\cite{UA1} UA1 discovered three new particles called $Z$, $W^+$ and $W^-$, predicted by the theory of weak interactions to be the mediators of the weak force. The mediators were produced by the annihilation between quarks and antiquarks and materialize in pairs of matter particles with a predictable topology. In fact, weak forces need three mediators to be explained, but they behave like photons with respect to Feynman diagrams. 

The existence of the mediators of strong forces, the {\em gluons}, was proven in 1979 looking at annihilations between electrons and positrons~\cite{gluons}.

Gravitons have not yet been discovered, because gravity is too weak to be experimentally investigated in colliders. There are other particles that are believed to exist and have not yet been observed. One of these particles is the Higgs boson: a neutral, massive particle responsible for the existence of the mass. Higgs bosons, according to the theory outlined above, can be produced at colliders just as photons. Unfortunately the probability to produce a Higgs boson is much lower than the one to produce photons; that's why it was not yet observed. Today's experiments try to observe it by producing a huge number of annihilations. Knowing the probability for its production one can estimate how many collisions are needed to be able to detect it, if exists. The search for the Higgs boson is currently one of the key points in the LHC scientific programme.

\section{Acknowledgements}
I am grateful to Prof. Guido Martinelli, Director of the SISSA (Trieste, Italy), who read the paper and provided some useful suggestions.

\section{Conclusion}
We introduced the technique of Feynman diagrams using very basic and popular concepts, explaining why particle physicists describe the interaction between matter particles as the exchange of mediators.


\vskip 1cm
\begin{thebibliography}{99}
\bibitem{wick} G.C. Wick, The Evaluation of the Collision Matrix, Phys. Rev. 80, 268 - 272 (1950).
\bibitem{feynman}R.P. Feynman, Space-time approach to quantum electrodinamics. Phys. Rev. 76, 769--789 (1949).
\bibitem{daniel}M. Daniel, "Particles, Feynman diagrams and all that", Phys. Educ. 41, 119.
\bibitem{dirac}P. A. M. Dirac, The quantum theory of the electron (1929)
\bibitem{anderson}C.D. Anderson, The Positive Electron, Phys. Rev. 43 (6): 491-Ð494 (1933).
\bibitem{SpS}UA1 collaboration, Experimental observation of events with large missing transverse energy accompanied by a jet or a photon(s) in  collisions at  $\sqrt{s} = 540$~GeV, Phys. Lett. B 139 (1984) 115--125
\bibitem{UA1}UA1 collaboration, Intermediate-mass dimuon events at the CERN  collider at $\sqrt{s} = 540$~GeV, Phys. Lett. B 155 (1985) 442--456
\bibitem{gluons}TASSO collaboration, Evidence for Planar Events in $e^+e^-$ Annihilation at High Energies, Phys. Lett. B 86, 243Ð-249 (1979).
\end{thebibliography}
\end{document}